\documentclass[aps,pra,twocolumn,showpacs]{revtex4-2}
\usepackage{epsfig,amssymb,amsfonts}
\usepackage{amsmath,amsfonts,amssymb}
\usepackage{graphicx}
\usepackage{caption}
\usepackage{xcolor}
\input{epsf}
\usepackage{epstopdf}
\usepackage{natbib}
\usepackage{braket}

\usepackage[labelformat=simple]{subcaption}
\makeatletter
\renewcommand\p@subfigure{\thefigure}
\makeatother

\begin{document}

\title{Fast chiral resolution with optimal control}
\author{Dionisis Stefanatos}
\email{dionisis@post.harvard.edu}
\author{Ioannis Thanopulos}
\author{Emmanuel Paspalakis}
\affiliation{Materials Science Department, School of Natural Sciences, University of Patras, Patras 26504, Greece}
\date{\today}
\begin{abstract}

In this work, we formulate the problem of achieving in minimum-time perfect chiral resolution with bounded control fields, as an optimal control problem on two non-interacting spins-$1/2$. We assume the same control bound for the two Raman fields (pump and Stokes) and a different bound for the field connecting directly the two lower-energy states. Using control theory, we show that the optimal fields can only take the boundary values or be zero, the latter corresponding to singular control. Subsequently, using numerical optimal control and intuitive arguments, we identify some three-stage symmetric optimal pulse-sequences, for relatively larger values of the ratio between the two control bounds, and analytically calculate the corresponding pulse timings as functions of this ratio. For smaller values of the bounds ratio, numerical optimal control indicates that the optimal pulse-sequence loses its symmetry and the number of stages increases in general. In all cases, the analytical or numerical optimal protocol achieves a faster perfect chiral resolution than other pulsed protocols, mainly because of the simultaneous action of the control fields. The present work is expected to be useful in the wide spectrum of applications across the natural sciences where enantiomer separation is a crucial task.

\end{abstract}

\maketitle

\section{Introduction}

The chiral asymmetry of molecules lies at the heart of many important applications in chemistry, biological, and pharmaceutical sciences, due to its importance for living organisms \cite{COMPTON02}.
Additionally, in physics, chiral molecules can be exploited for the emission and detection of chiral light, as well as for the control of molecular motion or electron spin and bulk charge transport \cite{Brandt17}.
They also provide an attractive experimental platform for fundamental symmetry tests \cite{Quack08,Erez23}.
For these reasons, the efficient separation of enantiomers with opposite left ($L$) and right (R) handedness, also known as chiral resolution, is a significant scientific endeavor.

\begin{figure}[t]
 \centering
\includegraphics[width=\linewidth]{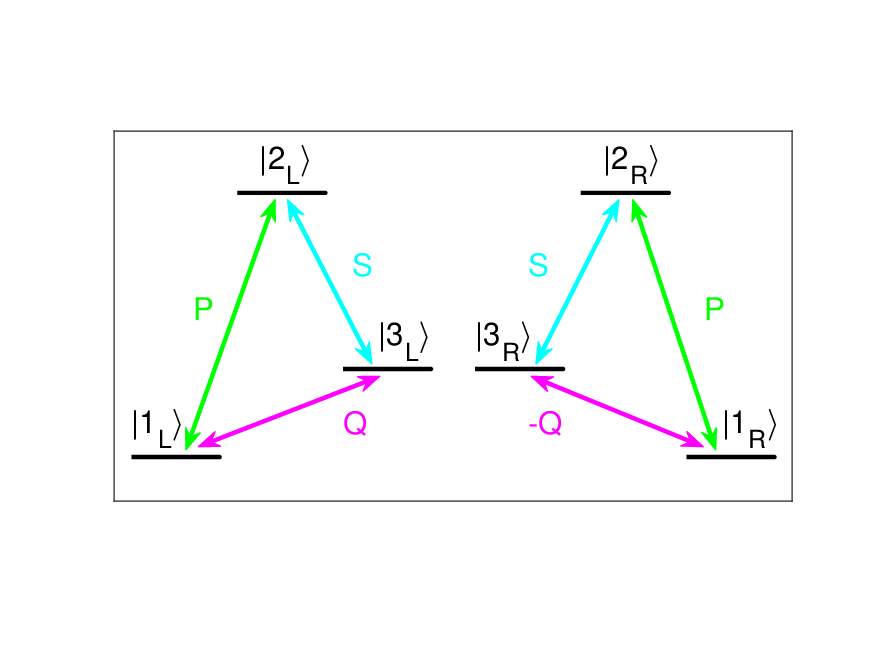}
\caption{Closed loop three-level system for the $L$- and $R$-molecules. The pump and Stokes Raman fields are common for both chiralities, while the $Q$-field has opposite sign.}
\label{fig:system}
\end{figure}

The enantiomers are modeled as closed-loop three-level systems, see Fig. \ref{fig:system}, where the two lower-energy states are connected to the excited state through the usual pump ($P$) and Stokes ($S$) Raman fields, while there is a third field ($Q$) connecting them directly, which closes the loop. The pump and Stokes fields are common for both $L$- and $R$-molecules, while the $Q$-field has a different sign for the two enantiomers. Both types of chiral molecules start from the same lower-energy state, and chiral resolution is achieved by applying appropriate control fields which drive one of the enantiomers to the other lower-energy state, while simultaneously suppressing the occupation of the same state for the other enantiomer. 

Several quantum control methods \cite{Stefanatos_2020} have been proposed to address this important task. Adiabatic passage methods \cite{Kral07,Vitanov17}, which are resilient against moderate experimental imperfections, have been suggested for chiral resolution in a series of works \cite{Kral01,Kral03,Thanopulos03,Gerbasi04,Thanopulos05}. To accelerate slow adiabatic dynamics, the method of shortcuts to adiabaticity \cite{Shortcuts19}, which does not follow the adiabatic trajectory at each instant but leads the system to the same final state, was proposed in Ref. \cite{Vitanov19} and also used in subsequent studies \cite{Wu19,Ding22,Cheng23a,Cheng23b,Xu23}. In parallel, various pulsed methods have been employed to properly design the control fields \cite{Ye19,Leibscher19,Torosov20a,Torosov20b,Wu20,Ayuso21,Guo22,Gong22,Leibscher22a,Leibscher22b,Erez23,Ye23,Cheng24,Leibscher24}.

In the present study, we express the problem of obtaining in minimum-time perfect chiral resolution with bounded control fields, as an optimal control problem on a system of two non-interacting spins-$1/2$. Note that quantum optimal control is a very active and productive area of research, see for example the recent reviews \cite{Boscain21,Koch22,SugnyReview}. We consider a common control bound for the pump and Stokes fields and a different bound for the $Q$-field connecting directly the two lower-energy levels. We first employ optimal control theory to show that the only values permitted for the optimal fields are the control bounds and zero, where the last one corresponds to singular control. Next, with the aid of numerical optimal control as well as some intuitive arguments, we single out three-stage symmetric pulse-sequences which appear to be optimal for relatively larger values of the ratio of the control bounds.
For these optimal sequences, we analytically compute the duration of the constituent pulses, as functions of the bounds ratio. For lower values of this ratio, numerical optimal control suggests that the optimal pulse-sequences are no longer symmetric and include more stages in general. In any case, the analytical or numerical optimal protocol attains a faster perfect chiral resolution than several pulsed protocols, primarily as a result of the synchronous activation of the control fields. The current study is anticipated to find application in various fields of science where enantiomer resolution plays a prominent role, such as coherent molecular dynamics \cite{Lee22,Leibscher24,Sun25}, molecular switches \cite{Thanopulos2009,Leeuwen2016}, photoelectron spectroscopy \cite{Lee24}, selective electron transfer \cite{Safari24}, creation of entangled states \cite{Kral2005}, and even food science \cite{Rivera20}, just to name  a few.

The article is structured in the following way. In Sec. \ref{sec:formulation} we formulate the optimal control problem of achieving chiral resolution in minimum time and in Sec. \ref{sec:analysis} we use optimal control theory to find the allowed values of the optimal control functions. We also discuss a solution with constant controls, emerging when the bounds on all fields are equal. In Secs. \ref{sec:greater} and \ref{sec:lower} we present optimal protocols for the cases where the bound on $Q$-control is respectively greater or smaller than the bound on the Raman fields. In Sec. \ref{sec:comparison} we compare the derived optimal protocols with other pulsed protocols, while Sec. \ref{sec:conclusion} concludes this study.

\section{Chiral resolution as an optimal control problem on two non-interacting spins}
\label{sec:formulation}

We consider the closed-loop three-level system for chiral molecules expressed by the following 
Schr\"odinger equation \cite{Vitanov19}
\begin{equation}
\label{LR_Hamiltonian}
i\left(
\begin{array}{c}
\dot{a}_{\pm} \\ 
\dot{b}_{\pm} \\ 
\dot{c}_{\pm}
\end{array}
\right)
=
\frac{1}{2}
\left(
\begin{array}{ccc}
0 & \Omega_p & \pm\Omega_qe^{i\phi}\\
\Omega_p^* & 0 & \Omega_s\\
\pm\Omega_qe^{-i\phi} & \Omega_s^* & 0
\end{array}
\right)
\left(
\begin{array}{c}
a_{\pm} \\ 
b_{\pm} \\ 
c_{\pm}
\end{array}
\right),
\end{equation}
where the plus sign corresponds to $L$-handed molecules and the minus sign to $R$-handed, while $a, b, c$ denote the probability amplitudes of states $\ket{1}, \ket{2}$ and $ \ket{3}$, respectively.
All three transitions can be simultaneously driven by the $P$-, $Q$- and $S$-fields, while the only difference for molecules with different handedness is the sign of the $Q$-field. Note that in Eq. (\ref{LR_Hamiltonian}) all detunings have been set to zero and the phase of the $Q$-field is taken to be $\phi=\pi/2$, as in Ref. \cite{Vitanov19}. In this work we consider real $\Omega_p, \Omega_q, \Omega_s$ except Sec. \ref{sec:comparison}, where we compare the derived optimal protocols with other pulsed protocols using different phases for the fields.

If we make the transformation
\begin{equation}
\label{transformation}
x_{\pm}=-c_{\pm}, \quad y_{\pm}=-ib_{\pm}, \quad z_{\pm}=a_{\pm},
\end{equation}
then the Bloch vectors 
$\vec{r}_\pm = (x_{\pm}, y_{\pm}, z_{\pm})^T$,
corresponding to the $L$- and $R$-molecules, obey the equations
\begin{equation}
\label{LR_spins}
\dot{\vec{r}}_{\pm} = \frac{1}{2}\vec{B}_{\pm}(t)\times \vec{r}_{\pm},
\end{equation}
where
\begin{equation}
\label{B}
\vec{B}_{\pm}(t) = \left[\Omega_p(t), \pm\Omega_q(t), \Omega_s(t)\right]^T.
\end{equation}


Both enantiomers start at $t=0$ from state $\ket{1}$, where $a_{\pm}(0)=1$ and $b_{\pm}(0)=c_{\pm}(0)=0$, so the common initial condition for the Bloch vectors is
\begin{equation}
\label{initial}
\vec{r}_\pm (0) = (0, 0, 1)^T.
\end{equation}
Chiral resolution is accomplished when, with the application of the appropriate fields, the molecules of one enantiomer are transferred to state $\ket{3}$, while those of the other are not.
The corresponding objective to be maximized at the final time $t=T$ using the control fields $\Omega_p, \Omega_q, \Omega_s$ is
\begin{equation}
\label{objective}
|x^2_+(T)-x^2_-(T)|=||c_+(T)|^2-|c_-(T)|^2|.
\end{equation}
To make the optimal control problem well-defined we impose the following 
constraints on the control functions
\begin{equation}
\label{constraints}
-\Omega_0\leq\Omega_{p, s}(t)\leq\Omega_0,\quad -\Omega_1\leq\Omega_q(t)\leq\Omega_1,
\end{equation}
where note that, driven by physical considerations, we treat the Raman fields on an equal footing using the same bound $\Omega_0$, while we use a different bound $\Omega_1$ for $\Omega_q$.

Observe that objective (\ref{objective}) attains its maximum possible value one when $|x_+(T)|=1$ and $|x_-(T)|=0$ ($|x_-(T)|=1$ and $|x_+(T)|=0$), where all the $L (R)$-molecules are excited, while the $R (L)$-molecules are not (perfect chiral resolution). In the rest of this work we concentrate on the problem of exciting only the $L$-molecules, the other case being equivalent. Achieving the maximum possible efficiency in minimum time is equivalent to finding optimal controls which simultaneously drive in minimum time the $\vec{r}_+$ vector on the $x$-axis and the $\vec{r}_-$ vector on the $yz$-meridian of the Bloch sphere. 
The terminal condition for the $\vec{r}_+$ vector is thus
\begin{equation}
\label{rplus_final}
\vec{r}_+(T) = (1, 0, 0)^T,
\end{equation}
where note that the negative $x$-axis case is equivalent and thus not considered  further,
while the $\vec{r}_-(T)$ should satisfy
\begin{equation}
\label{rm_final}
x_-(T)=0.
\end{equation}
In the paper we study this minimum-time optimal control problem.

Let $\vec{\lambda}_\pm = (\lambda_{\pm x}, \lambda_{\pm y}, \lambda_{\pm z})^T$ be the adjoint vectors corresponding to the state vectors $\vec{r}_\pm$. For minimizing the time $T=\int_0^T1dt$, the control Hamiltonian turns out to be \cite{Pontryagin62,Heinz12}
\begin{equation}
\label{Hc1}
H_c = -1 + \vec{\lambda}_+^T\cdot \dot{\vec{r}}_+ + \vec{\lambda}_-^T\cdot \dot{\vec{r}}_-.
\end{equation}
The state and adjoint vectors satisfy Hamilton's equations
\begin{subequations}
\begin{eqnarray}
\label{Hamilton_pair}
\dot{\vec{r}}_{\pm} &=& \frac{\partial H_c}{\partial\vec{\lambda}_{\pm}^T}, \label{rvec} \\
\dot{\vec{\lambda}}_{\pm}^T &=& -\frac{\partial H_c}{\partial\vec{r}_{\pm}}, \label{lvec}
\end{eqnarray}
\end{subequations}
where note that Eq. (\ref{rvec}) is equivalent to Eq. (\ref{LR_spins}), while Eq. (\ref{lvec}) leads to the same equation for the adjoint vectors, i.e.
\begin{equation}
\label{lambdavec}
\dot{\vec{\lambda}}_{\pm} = \frac{1}{2}\vec{B}_{\pm}(t)\times \vec{\lambda}_{\pm}.
\end{equation}

According to optimal control theory and specifically Pontryagin's Maximum Principle \cite{Pontryagin62,Heinz12,Boscain21,SugnyReview}, since the terminal $\vec{r}_-$ vector should lie on the $yz$-meridian, the adjoint vector $\vec{\lambda}_{-}$ should satisfy the corresponding transversality condition at the terminal point, which states that it should be orthogonal to the tangent vector of the terminal curve (the $yz$-meridian) and together should generate the tangent space of the ambient manifold (the Bloch sphere) at that point. This leads to the condition $\vec{\lambda}_-(T) \mathbin{\!/\mkern-5mu/\!} \hat{x}$, which gives
\begin{equation}
\label{transversality}
\lambda_{-y}(T) = \lambda_{-z}(T) = 0.
\end{equation}
Relations (\ref{rm_final}) and (\ref{transversality}) are the terminal conditions for the $\vec{r}_-, \vec{\lambda}_{-}$ pair which, along with the terminal condition (\ref{rplus_final}) for the $\vec{r}_+, \vec{\lambda}_{+}$ pair and the common initial condition (\ref{initial}), form a two-point boundary value problem for Eqs. (\ref{LR_spins}) and (\ref{lambdavec}).

In the following section we use optimal control theory to find the allowed values of the optimal pulses and also derive a constant control solution for perfect chiral resolution.

\section{Determination of optimal control values and a constant control solution}

\label{sec:analysis}

Before moving to discuss how the optimal fields $\vec{B}_{\pm}(t)$ are selected, it is necessary first to express the control Hamiltonian in a more compact way.
Following Ref. \cite{VanDamme18} we define the angular momentum vectors
\begin{equation}
\label{ang_mom}
\vec{L}_{\pm} = \vec{r}_{\pm}\times\vec{\lambda}_{\pm}, 
\end{equation}
and the corresponding sum and difference vectors
\begin{equation}
\label{sum_dif}
\vec{S}=\vec{L}_+ + \vec{L}_-, \quad \vec{D}=\vec{L}_+ -\vec{L}_-.
\end{equation}
Then, it is not hard to see that the control Hamiltonian can be expressed as
\begin{equation}
\label{Hc2}
H_c = -1 + \frac{1}{2}(S_x\Omega_p+D_y\Omega_q+S_z\Omega_s).
\end{equation}

In accordance with Maximum Principle \cite{Pontryagin62,Heinz12,Boscain21,SugnyReview}, the optimal controls $\Omega^*_p(t), \Omega^*_q(t), \Omega^*_s(t)$ are selected to maximize the control Hamiltonian for almost all times (except possibly a measure zero set). Since the control Hamiltonian is linear in the bounded control variables $\Omega_p, \Omega_q, \Omega_s$, the optimal pulse-sequence is determined by the switching functions $S_x, D_y, S_z$ multiplying these controls. For example, $\Omega^*_q(t)=\Omega_1$ when $D_y>0$ while $\Omega^*_q(t)=-\Omega_1$ for $D_y<0$. In optimal control terminology, when the control function takes the boundary values of the allowed interval, we refer to them as bang controls. The Maximum Principle provides no information about the optimal control for finite time intervals where the switching function is zero. In such cases the optimal control can take values from the interior of the allowed interval, is called singular and is determined from the requirements  $D_y=\dot{D}_y=\ddot{D}_y=\ldots=0$ for $\Omega_q$ and analogous for the others. It is obvious from the above discussion that, to find the optimal control functions, we need to track the time evolution of the switching functions $S_x, D_y, S_z$. Using Eqs. (\ref{LR_spins}), (\ref{lambdavec}) and the Jacobi identity, it is not hard to verify that the angular momentum vectors satisfy also the equation 
\begin{equation}
\dot{\vec{L}}_{\pm} = \frac{1}{2}\vec{B}(t)\times\vec{L}_{\pm},
\end{equation}
and from this 
we find that the components of $\vec{S}=(S_x, S_y, S_z)^T$ and $\vec{D}=(D_x, D_y, D_z)^T$ obey the equations
\begin{subequations}
\label{SD}
\begin{eqnarray}
\dot{S}_x &=& -\frac{1}{2}(\Omega_s S_y-\Omega_q D_z), \\
\dot{S}_y &=& \frac{1}{2}(\Omega_s S_x-\Omega_p S_z), \\
\dot{S}_z &=& -\frac{1}{2}(\Omega_q D_x-\Omega_p S_y), \\
\dot{D}_x &=& -\frac{1}{2}(\Omega_s D_y-\Omega_q S_z), \\
\dot{D}_y &=& \frac{1}{2}(\Omega_s D_x-\Omega_p D_z), \\
\dot{D}_z &=& -\frac{1}{2}(\Omega_q S_x-\Omega_p D_y).
\end{eqnarray}
\end{subequations}

Now suppose that $D_y(t)=0$ on some finite time interval, while the other two switching functions are nonzero on this interval, thus $\Omega_p, \Omega_s$ are constant and equal to one of the boundary values. Then, on the same interval it is also
\begin{equation}
\label{dotD_y}
\dot{D}_y = \frac{1}{2}(\Omega_s D_x-\Omega_p D_z) = 0 
\end{equation}
and
\begin{equation}
\label{ddotD_y}
\ddot{D}_y = \frac{1}{2}(\Omega_s \dot{D}_x-\Omega_p \dot{D}_z) = 0,
\end{equation}
where we have used the constancy of $\Omega_p, \Omega_s$.
Using Eqs. (\ref{SD}) in Eq. (\ref{ddotD_y}) we get the relation
\begin{equation}
\label{find_singular}
\Omega_q (S_x\Omega_p + S_z\Omega_s) = 0 
\end{equation}
For a control system without explicit time-dependence but only through the control functions, like the one here, Maximum Principle assures that the control Hamiltonian is constant. Additionally, for a minimum time problem this constant is actually zero. From Eq. (\ref{Hc2}), using that $H_c = 0$ and also $D_y = 0$ on the singular arc, we obtain $S_x\Omega_p+S_z\Omega_s = 2$ which, in combination with Eq. (\ref{find_singular}) leads to the singular control value
\begin{equation}
\label{qsingular}
\Omega_q(t) = 0.
\end{equation}

If we consider that $S_z=0$ on some finite time interval while the other two switching functions are not then, working similarly, we obtain the relation
\begin{equation}
\Omega_s (S_x\Omega_p + D_y\Omega_q) = 0 
\end{equation}
which, using also that $H_c=0$, leads to
\begin{equation}
\label{ssingular}
\Omega_s(t) = 0.
\end{equation}
Similarly, $S_x=0$ leads to
\begin{equation}
\Omega_p (D_y\Omega_q+S_z\Omega_s) = 0 
\end{equation}
which, along with $H_c=0$ gives
\begin{equation}
\label{psingular}
\Omega_p(t) = 0.
\end{equation}
If two of the switching functions are simultaneously zero on some finite interval while the third is not, then we can show with similar arguments that the corresponding singular controls are both zero. 
All three switching functions cannot vanish simultaneously on an interval, as this leads to $H_c=-1\neq 0$.

The conclusion from the above analysis is that the optimal control functions can assume only three values, the two boundary values corresponding to bang controls and zero corresponding to singular control.

Motivated by the above findings and to gain some intuition, we first investigate if there is a constant control solution which can achieve perfect chiral resolution.
To facilitate the computations here and in the following sections, we will use instead of the Bloch vectors
the corresponding two-level states 
\begin{equation}
\ket{\Psi_{\pm}} = 
\left(
    \begin{array}{c}
        A_{\pm}\\
        B_{\pm}
    \end{array} 
\right), 
\end{equation}
where
\begin{subequations}
\label{qubit_transf}
    \begin{eqnarray}
     z_{\pm} &=& |A_{\pm}|^2-|B_{\pm}|^2,\\
     y_{\pm} &=& 2\Im{(A^*_{\pm}B_{\pm}}),\\
     x_{\pm} &=& 2\Re{(A^*_{\pm}B_{\pm})},
    \end{eqnarray}
\end{subequations}
and $\Re,\Im$ denote real and imaginary parts, respectively.
They obey the equations
\begin{equation}
\label{Eq_2}
i \frac{d\ket{\Psi_{\pm}}}{dt} =\hat{H}_{\pm}(t)\ket{\Psi_{\pm}}
\end{equation}
with
\begin{equation}
\label{H}
\hat{H}_{\pm}(t)
= \frac{\Omega_p(t)}{4}\hat{\sigma}_x\pm\frac{\Omega_q(t)}{4} \hat{\sigma}_y + \frac{\Omega_s(t)}{4}\hat{\sigma}_z,
\end{equation}
$\sigma_i, i=x, y, z$ being the Pauli matrices.

\begin{figure}[t]
 \centering
 \begin{subfigure}[b]{0.4\textwidth}
    \centering\caption{}\includegraphics[width=\linewidth]{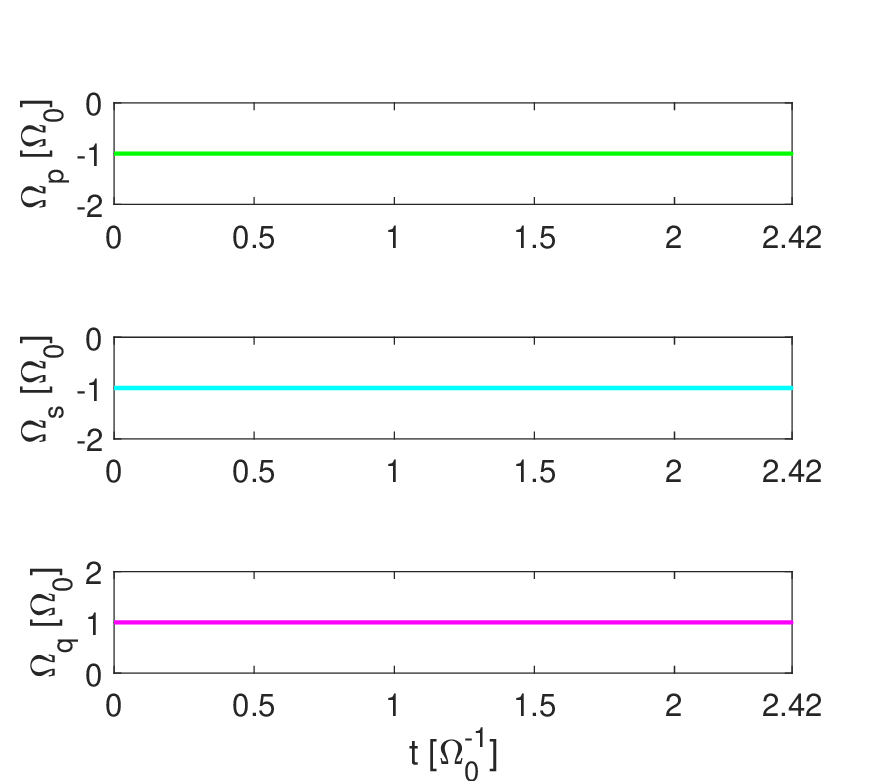}\label{fig:W1c}
\end{subfigure}%
\hspace{.2cm}
\begin{subfigure}[b]{0.4\textwidth}
    \centering\caption{}\includegraphics[width=\linewidth]{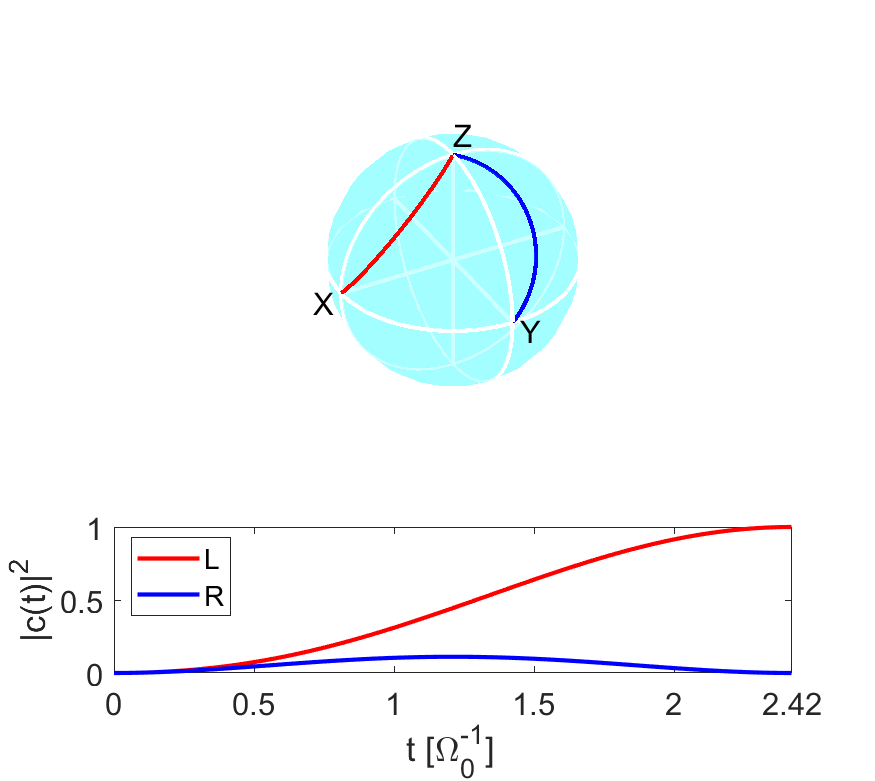}\label{fig:W1}
\end{subfigure}
\caption{Case $\Omega_1/\Omega_0=1$, (a) optimal controls, (b) optimal trajectories on the Bloch sphere (upper panel) and time evolution of level $\ket{3}$ populations (lower panel) for the $L$- (red) and $R$- (blue) molecules.}
\label{fig:1}
\end{figure}

We thus consider constant control functions $\Omega_p(t) =\Omega_p$, $\Omega_q(t) =\Omega_q$, $\Omega_s(t) =\Omega_s$ 
applied for the whole time interval $T$. If we set
\begin{equation}
\Omega=\sqrt{\Omega_p^2+\Omega_s^2+\Omega_q^2},
\end{equation}
then the corresponding propagator is
\begin{eqnarray}
\label{const_prop}
U_{\pm T} &=& e^{-\frac{i}{4}\Omega T(n_x\sigma_x\pm n_y\sigma_y+n_z\sigma_z)} \nonumber \\
    &=& I \cos{\frac{\Omega T}{4}}-i(n_x\sigma_x\pm n_y\sigma_y+n_z\sigma_z)\sin{\frac{\Omega T}{4}},
\end{eqnarray}
where
\begin{equation}
n_x = \frac{\Omega_p}{\Omega},\quad n_y = \frac{\Omega_q}{\Omega},\quad n_z = \frac{\Omega_s}{\Omega},
\end{equation}
while we remind that the $\pm$ sign corresponds to the $L$- and $R$-handed molecules.
The target propagator for perfect chiral resolution is
\begin{equation}
\label{target}
U_{\pm tar}=\frac{1}{2}(I+i\sigma_x\mp i\sigma_y+i\sigma_z).
\end{equation}
Comparing Eqs. (\ref{const_prop}) and (\ref{target}) we find $\cos(\Omega T/4)=1/2$, thus the shortest possible duration is $T=4\pi/(3\Omega)$, while additionally $n_x=-n_y=n_z=-1/\sqrt{3}$. This last relation implies that we should choose $\Omega_1=\Omega_0$ and $\Omega_p=\Omega_s=-\Omega_q=-\Omega_0$, thus $\Omega=\Omega_0\sqrt{3}$ and the minimum duration for perfect resolution with constant fields is
\begin{equation}
\label{const}
T = \frac{4\pi}{3\sqrt{3}}\frac{1}{\Omega_0}.
\end{equation}
The constant controls  are displayed in Fig. \ref{fig:W1c}, while the corresponding trajectories of the $L$- and $R$-vectors on the Bloch sphere are plotted in the upper panel of Fig. \ref{fig:W1}, with red and blue color, respectively. The time evolution of state $\ket{3}$ population for the two enantiomers is shown in the lower panel of the same figure.

The finding that perfect chiral resolution with constant controls requires the control bounds to be equal, $\Omega_1=\Omega_0$, leads us to investigate separately the cases where $\Omega_1>\Omega_0$ and $\Omega_1<\Omega_0$.

\section{Optimal solution for $\Omega_1>\Omega_0$}

\label{sec:greater}

\begin{figure}[t]
 \centering
 \begin{subfigure}[b]{0.4\textwidth}
    \centering\caption{}\includegraphics[width=\linewidth]{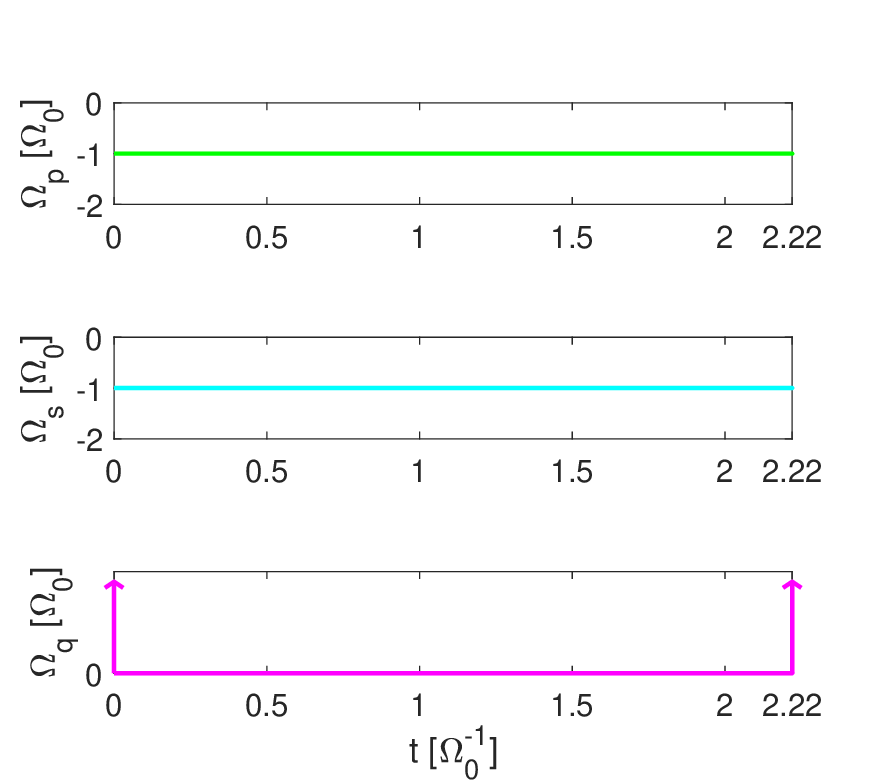}\label{fig:Winfc}
\end{subfigure}%
\hspace{.2cm}
\begin{subfigure}[b]{0.4\textwidth}
    \centering\caption{}\includegraphics[width=\linewidth]{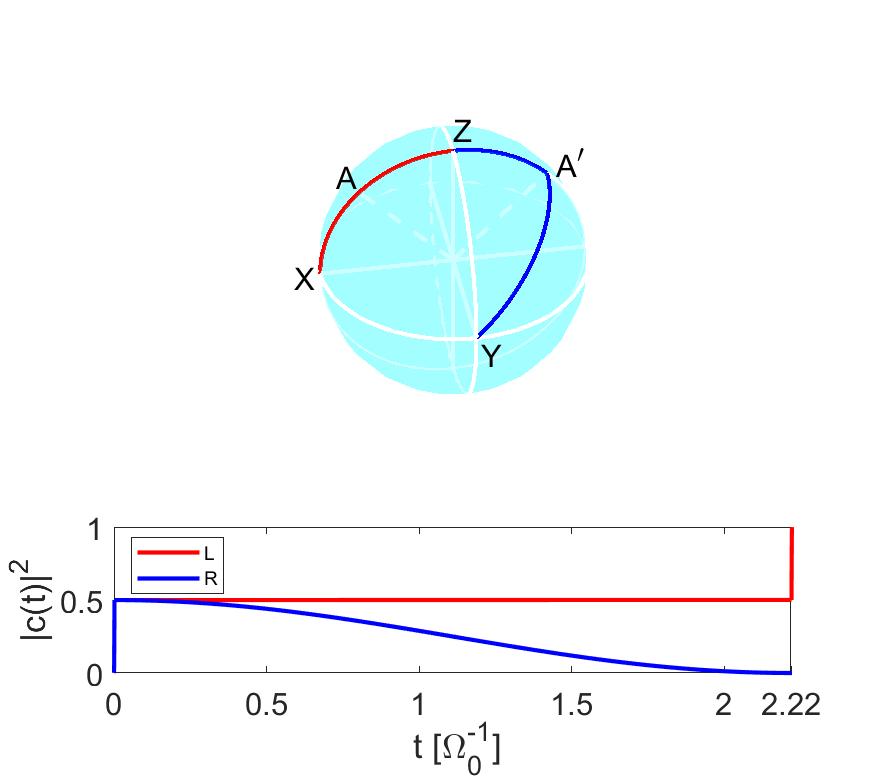}\label{fig:Winf}
\end{subfigure}
\caption{Case $\Omega_1/\Omega_0\rightarrow\infty$, (a) optimal controls, (b) optimal trajectories on the Bloch sphere (upper panel) and time evolution of level $\ket{3}$ populations (lower panel) for the $L$- (red) and $R$- (blue) molecules.}
\label{fig:inf}
\end{figure}

In order to develop some intuition using the model of two non-interacting spins, we start with the situation $\Omega_1\gg\Omega_0$, which permits instantaneous delta pulses in $\Omega_q$. For $\Omega_q(t)$ we consider a pulse-sequence consisting of two delta pulses at the beginning and end with area $\Omega_1\tau=\pi/2$, where $\tau\rightarrow 0$ as $\Omega_1\rightarrow \infty$, while $\Omega_q(t)=0$ in the rest of the time interval $0<t<T$. The other two control fields are set to the minimum value
$\Omega_p(t)=\Omega_s(t)=-\Omega_0$ for the whole interval $0\leq t\leq T$. The controls are shown in Fig. \ref{fig:Winfc}, where the initial and final delta pulses in $\Omega_q$ are schematically represented by arrows. At the beginning and end the delta pulses dominate so the corresponding propagators are 
\begin{equation}
U_{\pm 1}=U_{\pm 3}=e^{\mp i\frac{\Omega_1\tau}{4}\sigma_y}=e^{\mp i\frac{\pi}{8}\sigma_y},
\end{equation}
where the $\pm$ sign in the propagators corresponds to the $L$- and $R$-handed molecules,
while for $0<t<T$
\begin{equation}
\label{U2}
U_2=e^{i\frac{\Omega_0T}{4}(\sigma_x+\sigma_z)}=e^{-i\frac{\Omega_0T\sqrt{2}}{4}(\hat{n}\cdot\hat{\sigma})}
   \end{equation}
with 
\begin{equation}
\label{n}
\hat{n} = -\frac{1}{\sqrt{2}}\hat{x}-\frac{1}{\sqrt{2}}\hat{z}.
\end{equation}

We next move to find the pulse-sequence duration $T$ so the total propagator $U_{\pm T}=U_{\pm 3}U_2U_{\pm 1}$ reproduces the target propagator $U_{\pm tar}$.
We can visualize the above pulse-sequence on the Bloch sphere shown in the upper panel of Fig. \ref{fig:Winf}, where note that because of the $1/2$ factor in Eq. (\ref{LR_spins}) the $\pi/2$ pulses actually result in $\pi/4$ rotations. The first delta pulse around $y$-axis brings the $L$-vector to point $A$, in the middle of the $ZX$ arc, and the $R$-vector to the symmetric point $A'$, in the middle of the $Z(-X)$ arc. After this first rotation, the $L$-vector becomes co-linear with the axis $\hat{n}$ given in Eq. (\ref{n}), which is the rotation axis corresponding to the constant finite pulses applied in the interval $0<t<T$, while the $R$-vector lies on a plane perpendicular to this axis. Thus, the second rotation does not affect the $L$-vector, while it should bring the $R$-vector on the $y$-axis, along the $A'Y$ arc. The final delta pulse around $y$-axis does not affect the $R$-vector, while it rotates the $L$-vector another $\pi/4$ angle to the $x$-axis, along the $AX$ arc. From the above analysis becomes clear that to achieve a perfect chiral resolution, during the interval $0<t<T$ the $R$-vector should be rotated by a $\pi/2$ angle, from $A'$ to $Y$. From Eq. (\ref{U2}) this requirement is translated to the condition $\Omega_0T\sqrt{2}/2=\pi/2$, leading to
\begin{equation}
\label{glimit}
T = \frac{\pi}{\sqrt{2}}\frac{1}{\Omega_0}.
\end{equation}
It is not hard to verify algebraically that the total propagator satisfies $U_{\pm T}=U_{\pm 3}U_2U_{\pm 1}=U_{\pm tar}$. 
The time evolution of state $\ket{3}$ population for the two enantiomers is shown in the lower panel of Fig. \ref{fig:Winf}, with red for $L$- and blue for $R$-molecules. Note the abrupt changes at the beginning and end due to the delta pulses. We point that in a real molecular system such strong pulses may induce other effects, for example involve more excited states in the transfers. The goal of the above mathematical analysis was to develop some intuition regarding the optimal pulse sequence and derive the limiting duration (\ref{glimit}), which is a lower bound of the minimum necessary time for chiral resolution under system (\ref{LR_Hamiltonian}) when $\Omega_1$ is finite.

\begin{figure}[t]
 \centering
 \begin{subfigure}[b]{0.4\textwidth}
    \centering\caption{}\includegraphics[width=\linewidth]{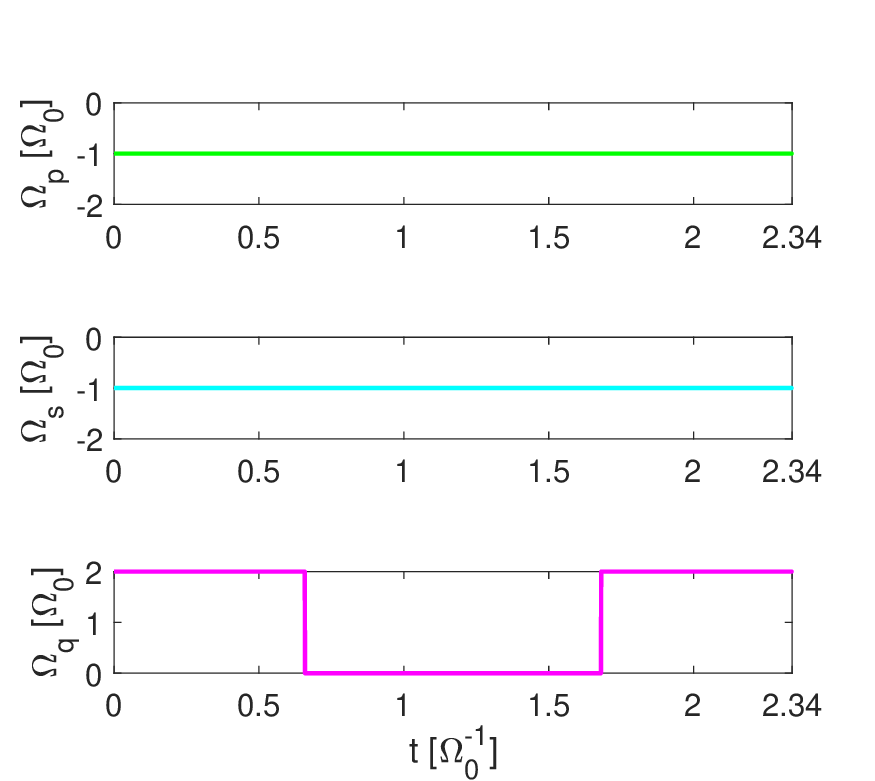}\label{fig:W2c}
\end{subfigure}%
\hspace{.2cm}
\begin{subfigure}[b]{0.4\textwidth}
    \centering\caption{}\includegraphics[width=\linewidth]{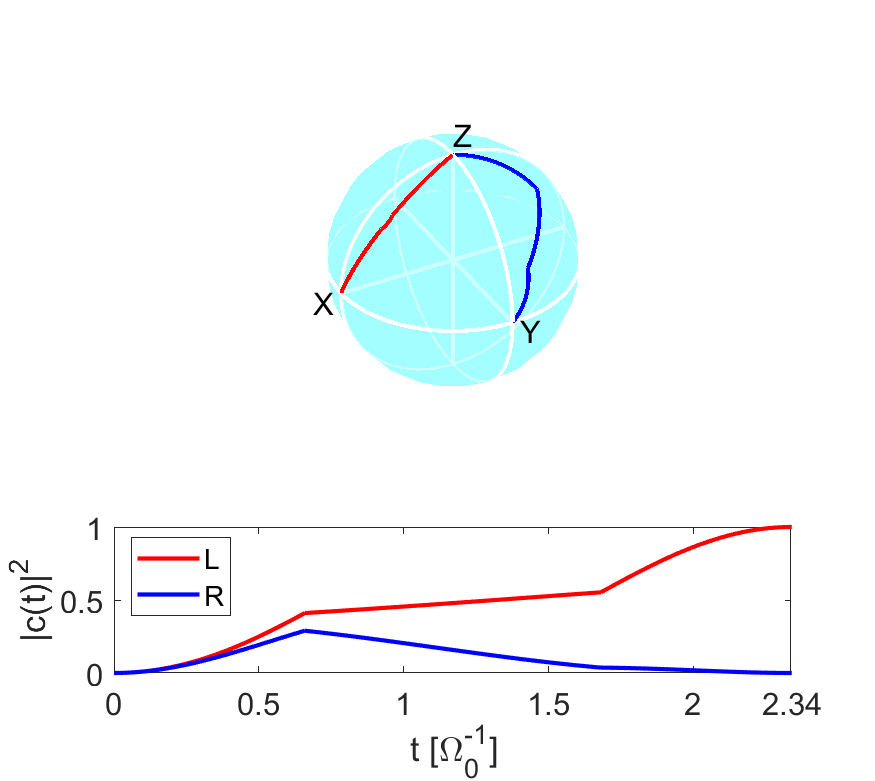}\label{fig:W2}
\end{subfigure}
\caption{Case $\Omega_1/\Omega_0=2$, (a) optimal controls, (b) optimal trajectories on the Bloch sphere (upper panel) and time evolution of level $\ket{3}$ populations (lower panel) for the $L$- (red) and $R$- (blue) molecules.}
\label{fig:2}
\end{figure}

For finite $\Omega_1>\Omega_0$ such that model (\ref{LR_Hamiltonian}) remains valid and no higher states are excited, the initial and final stages have a finite duration $\tau$, while the duration of the second stage occupies a smaller portion of the total duration and becomes $T-2\tau$. Extensive use of the optimal control solver BOCOP \cite{bocop} indicates that in this range of parameters the aforementioned pulse-sequence is the optimal solution, we thus move to find the durations $T, \tau$ for which perfect chiral resolution is achieved, for fixed $\Omega_1$. During the initial and final stages we have $\Omega_q(t)=\Omega_1$ and $\Omega_p(t)=\Omega_s(t)=-\Omega_0$, thus the corresponding propagators become
\begin{equation}
\label{finiteU13}
U_{\pm 1}=U_{\pm 3} = e^{-i\frac{\Omega_0\tau\sqrt{2+s^2}}{4}(\hat{n}_{\pm}\cdot\hat{\sigma})},
\end{equation}
where 
\begin{equation}
\label{s}
s=\frac{\Omega_1}{\Omega_0}>1, 
\end{equation}
\begin{equation}
\label{npm}
\hat{n}_{\pm} = -\frac{1}{\sqrt{s^2+2}}\hat{x}\pm\frac{s}{\sqrt{s^2+2}}\hat{y}-\frac{1}{\sqrt{s^2+2}}\hat{z},
\end{equation}
and the $\pm$ sign corresponds to the $L$- and $R$-molecules, respectively.
During the intermediate pulse it is $\Omega_q(t)=0$ and $\Omega_p(t)=\Omega_s(t)=-\Omega_0$, and the corresponding propagator is given by Eq. (\ref{U2}), with $T$ replaced by $T-2\tau$.
Putting all these together we find the total propagator to be
\begin{eqnarray}
\label{finonetot}
U_{\pm T} & = & U_{\pm 3}U_2U_{\pm 1} \nonumber \\
    & = & \left[\cos{\phi_1}\cos{\phi_2}-\sin{\phi_1}\sin{\phi_2}(\hat{n}\cdot\hat{n}_{\pm})\right]I \nonumber \\
    & + & i\left[\sin{\phi_2}(1-\cos{\phi_1})(\hat{n}\cdot\hat{n}_{\pm})-\sin{\phi_1}\cos{\phi_2}\right](\hat{n}_{\pm}\cdot\hat{\sigma}) \nonumber \\
    & - & i\sin{\phi_2}(\hat{n}\cdot\hat{\sigma}),
\end{eqnarray}
where
\begin{equation}
\label{phi}
\phi_1=\frac{\Omega_0\tau\sqrt{s^2+2}}{2}, \quad \phi_2=\frac{\Omega_0(T-2\tau)\sqrt{2}}{4}.
\end{equation}
In the derivation of Eq. (\ref{finonetot}) we have used the relations
\begin{subequations}
\begin{eqnarray}
(\hat{n}\cdot\hat{\sigma})(\hat{n}_{\pm}\cdot\hat{\sigma}) &=& \hat{n}\cdot\hat{n}_{\pm}+i\hat{\sigma}\cdot(\hat{n}\times\hat{n}_{\pm}) \\
\hat{n}_{\pm}\times (\hat{n}\times\hat{n}_{\pm}) &=& \hat{n}-(\hat{n}\cdot\hat{n}_{\pm})\hat{n}_{\pm}
\end{eqnarray}
\end{subequations}

To find the durations $T, \tau$, we equate the coefficients of $I, \sigma_y$ in Eq. (\ref{finonetot}) to the corresponding ones from Eq. (\ref{target}) for the target propagator and get the following relations
\begin{eqnarray}
n_x\sqrt{2}\sin{\phi_2}\sin{\phi_1} + \cos{\phi_2}\cos{\phi_1} &=& \frac{1}{2}, \label{linear_phi} \\
\cos{\phi_2}\sin{\phi_1} - n_x\sqrt{2}\sin{\phi_2}\cos{\phi_1} &=& \frac{1}{2n_y} - n_x\sqrt{2}\sin{\phi_2}, \nonumber
\end{eqnarray}
where we have set for compactness
\begin{equation}
\label{ncomp}
n_x=-\frac{1}{\sqrt{s^2+2}}, \quad n_y=\frac{s}{\sqrt{s^2+2}}.
\end{equation}
Eqs. (\ref{linear_phi}) form a linear system for $\sin{\phi_1}, \cos{\phi_1}$ which can be easily solved to give
\begin{equation}
\label{phi12}
\sin{\phi_1} = \frac{\Delta_s}{\Delta}, \quad \cos{\phi_1} = \frac{\Delta_c}{\Delta},
\end{equation}
where 
\begin{subequations}
\label{D}
\begin{eqnarray}
\Delta &=& 2n_x^2\sin^2{\phi_2}+\cos^2{\phi_2}, \\
\Delta_s &=& \frac{1}{\sqrt{2}}n_x\sin{\phi_2} + \cos{\phi_2}\left(\frac{1}{2n_y}-n_x\sqrt{2}\sin{\phi_2}\right), \\
\Delta_c &=& \frac{1}{2}\cos{\phi_2} - n_x\sqrt{2}\sin{\phi_2}\left(\frac{1}{2n_y}-n_x\sqrt{2}\sin{\phi_2}\right).
\end{eqnarray}
\end{subequations}

Using expressions (\ref{D}) in the identity $\sin^2{\phi_1}+\cos^2{\phi_1}=1$, we get after some manipulation that
\begin{equation}
\label{phi2eq}
\sin^2{\phi_2}+\frac{\sqrt{2}}{s}\sin{\phi_2}-\frac{1}{2}\left(1-\frac{1}{s^2}\right) = 0,
\end{equation}
where note that we have also used the relations (\ref{ncomp}).
The solution of Eq. (\ref{phi2eq}) giving the shorter duration is
\begin{equation}
\sin{\phi_2}=\frac{1}{\sqrt{2}}\left(1-\frac{1}{s}\right),
\end{equation}
from which and Eq. (\ref{phi}) we find the duration of the intermediate segment
\begin{equation}
\label{t2}
T-2\tau = \frac{2\sqrt{2}}{\Omega_0}\arcsin{\left[\frac{1}{\sqrt{2}}\left(1-\frac{1}{s}\right)\right]}.
\end{equation}
The most efficient way to obtain angle $\phi_1$ is to use Eq. (\ref{linear_phi}) to derive a quadratic equation for $\tan{(\phi_1/2)}$, whose acceptable (positive) solution turns out to be
\begin{equation}
\label{tanphi1}
\tan{\frac{\phi_1}{2}} = \frac{\sqrt{s^2+2}}{{s+\sqrt{2(s^2+2s-1)}}},
\end{equation}
leading to the following duration for the initial and final segments
\begin{equation}
\label{tau}
\tau = \frac{4}{\Omega_0}\frac{1}{\sqrt{s^2+2}}\arctan{\left[\frac{\sqrt{s^2+2}}{{s+\sqrt{2(s^2+2s-1)}}}\right]}.
\end{equation}
Using Eqs. (\ref{t2}) and (\ref{tau}) for $s=1$ ($\Omega_1=\Omega_0$) we find that $T=2\tau$ (the intermediate stage with $\Omega_q=0$ disappears) and $T$ attains the value given in Eq. (\ref{const}), 
while in the limit $s\rightarrow\infty$ it is $\tau\rightarrow 0$ and $T$ recovers the limiting value (\ref{glimit}). 
In Fig. \ref{fig:W2c} we display the optimal pulse-sequence for $s=\Omega_1/\Omega_0=2$, while in Fig. \ref{fig:W2} the corresponding trajectories of the Bloch vectors (upper panel) and the time evolution of state $\ket{3}$ populations for the $L$- (red) and $R$- (blue) molecules (lower panel).

\section{Optimal solution for $\Omega_1<\Omega_0$}

\label{sec:lower}

For cases with $s=\Omega_1/\Omega_0<1$, the optimal control solver BOCOP indicates that the efficiency is maximized with the $L$-vector arriving at the final time on the $x$-axis as before, while the $R$-vector ends up on the $yz$-meridian. This is somehow intuitively expected since now the maximum value of $\Omega_q$ is shorter than that of the other two fields, thus it cannot bring the two vectors on the $x$- and $y$-axes as before. We also observe that for larger values of $s$, up to $s\approx 0.86$, the optimal solution is composed of three stages, where the first and last have the same duration $\tau$ and also present some kind of symmetry. Specifically, $\Omega_q(t)=\Omega_1$ throughout the whole interval $T$, while the other two controls vary through the stages as follows. During the stage $0\leq t \leq \tau $ it is $\Omega_p(t)=-\Omega_0, \Omega_s(t)=0$, in the second stage $\tau < t \leq T-\tau$ it is $\Omega_p(t)=-\Omega_0, \Omega_s(t)=-\Omega_0$, while in the last stage $T-\tau < t \leq T$ it is $\Omega_p(t)=0, \Omega_s(t)=-\Omega_0$. In the following we derive analytical expressions for $T, \tau$ as functions of $s$, as we did in the previous case.

\begin{figure}[t]
 \centering
 \begin{subfigure}[b]{0.4\textwidth}
    \centering\caption{}\includegraphics[width=\linewidth]{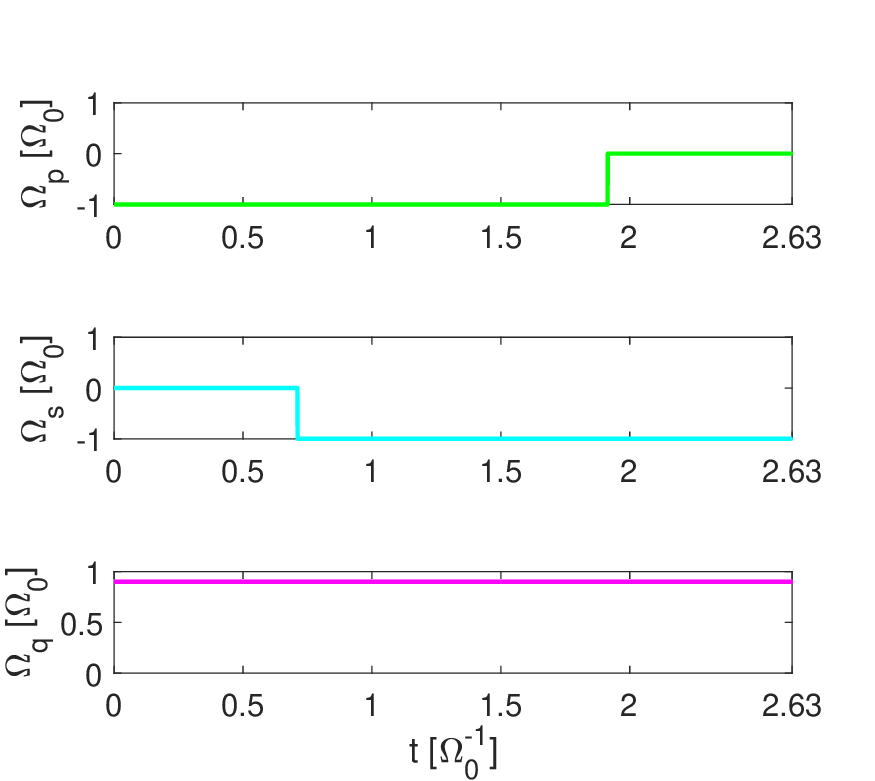}\label{fig:W09c}
\end{subfigure}%
\hspace{.2cm}
\begin{subfigure}[b]{0.4\textwidth}
    \centering\caption{}\includegraphics[width=\linewidth]{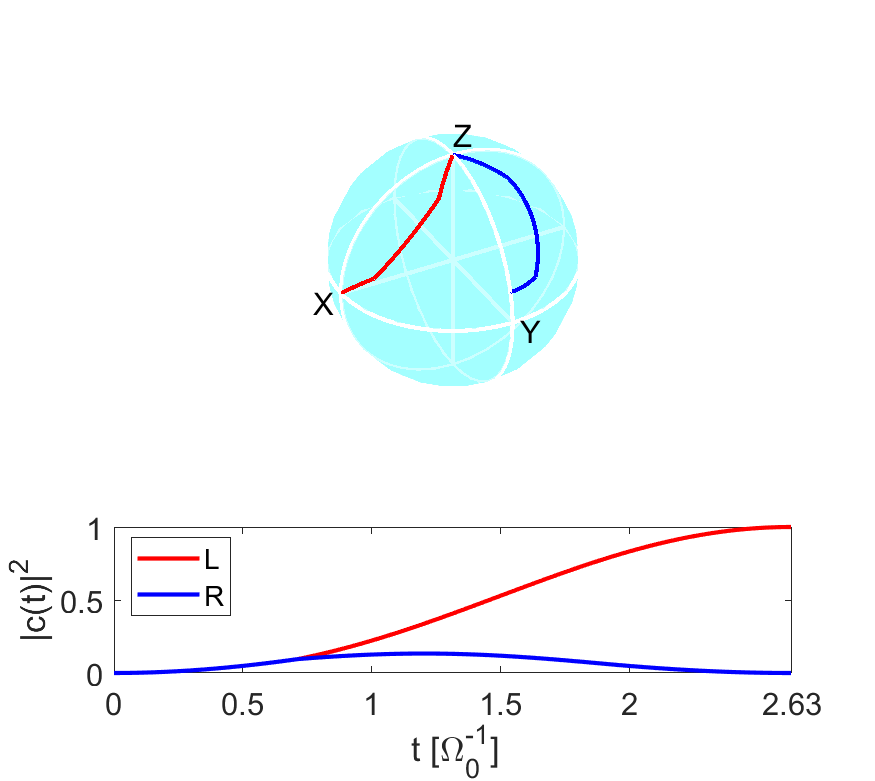}\label{fig:W09}
\end{subfigure}
\caption{Case $\Omega_1/\Omega_0=0.9$, (a) optimal controls, (b) optimal trajectories on the Bloch sphere (upper panel) and time evolution of level $\ket{3}$ populations (lower panel) for the $L$- (red) and $R$- (blue) molecules.}
\label{fig:09}
\end{figure}

The rotation axes during the initial and final stages are
\begin{subequations}
\label{n1n3}
\begin{eqnarray}
\hat{n}_{\pm 1} &=& w(n_x\hat{x}\pm n_y\hat{y}), \\
\hat{n}_{\pm 3} &=& w(\pm n_y\hat{y}+n_x\hat{z}),
\end{eqnarray}
\end{subequations}
where 
\begin{equation}
\label{r}
w=\sqrt{\frac{s^2+2}{s^2+1}}
\end{equation}
is a normalization constant assuring that $\hat{n}_1, \hat{n}_3$ are unit vectors. During the second stage the rotation axis is $\hat{n}_{\pm}$ given in Eq. (\ref{npm}).
The corresponding propagators are
\begin{subequations}
\label{smallU}
\begin{eqnarray}
U_{\pm 1,3} &=& e^{-i\phi_1(\hat{n}_{\pm 1,3}\cdot\hat{\sigma})}, \\
U_{\pm 2} &=& e^{-i\phi_2(\hat{n}_{\pm}\cdot\hat{\sigma})},
\end{eqnarray}
\end{subequations}
where 
\begin{equation}
\label{phiprime}
\phi_1=\frac{\Omega_0\tau\sqrt{s^2+1}}{4}, \quad \phi_2=\frac{\Omega_0(T-2\tau)\sqrt{s^2+2}}{4}.
\end{equation}

\begin{figure}[t]
 \centering
\includegraphics[width=\linewidth]{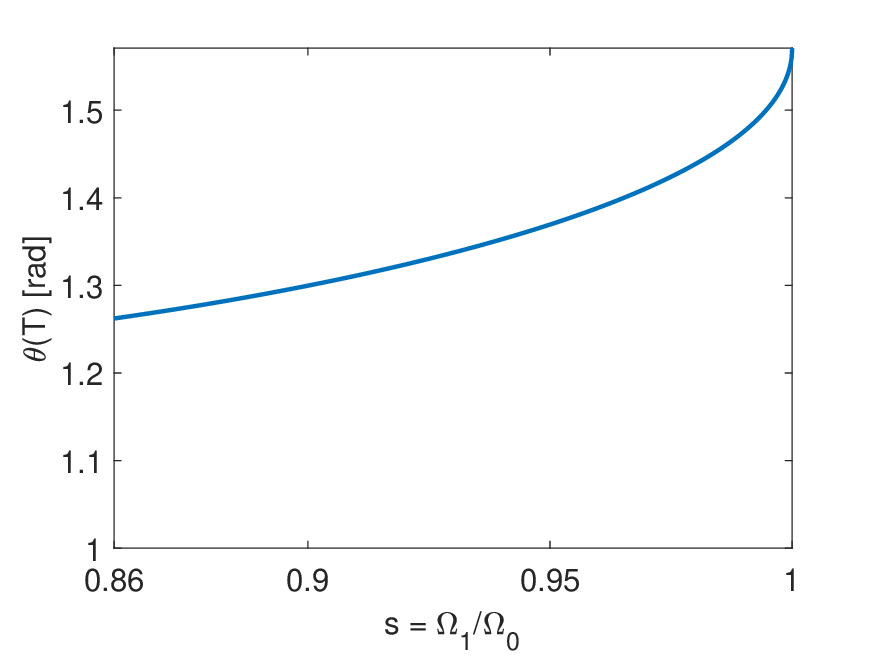}
\caption{Polar angle of the terminal point of the $R$-trajectory on the $yz$-meridian as a function of the ratio of the maximum control amplitudes.}
\label{fig:polar}
\end{figure}

\begin{figure}[t]
 \centering
 \begin{subfigure}[b]{0.4\textwidth}
    \centering\caption{}\includegraphics[width=\linewidth]{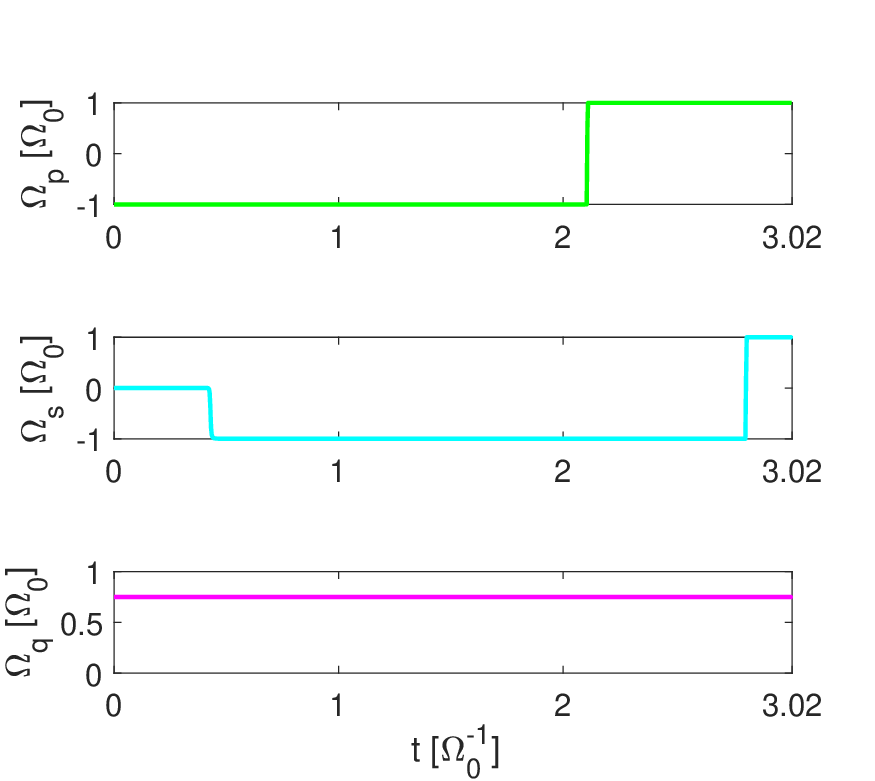}\label{fig:W075c}
\end{subfigure}%
\hspace{.2cm}
\begin{subfigure}[b]{0.4\textwidth}
    \centering\caption{}\includegraphics[width=\linewidth]{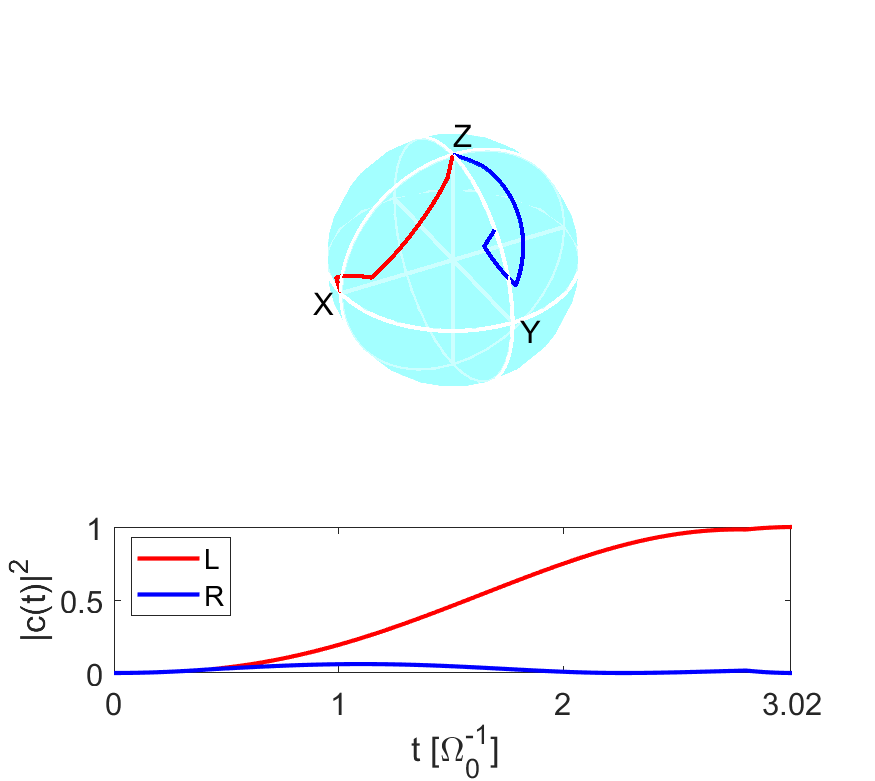}\label{fig:W075}
\end{subfigure}
\caption{Case $\Omega_1/\Omega_0=0.75$, (a) optimal controls, (b) optimal trajectories on the Bloch sphere (upper panel) and time evolution of level $\ket{3}$ populations (lower panel) for the $L$- (red) and $R$- (blue) molecules.}
\label{fig:075}
\end{figure}

The first relation connecting $T, \tau$ will be obtained from the condition
\begin{equation}
\label{xone}
x_+(T)=2\Re{\{A^*_+(T)B_+(T)\}}=1.
\end{equation}
It is not hard to verify that this condition implies for the components of the final state of the $L$-system that
\begin{equation}
\label{equality}
A_+(T)=B_+(T).
\end{equation}
We find these components from the following propagation equation 
\begin{equation}
U_{+3}^\dagger
\left[
    \begin{array}{c}
        A_+(T) \\
        B_+(T)
    \end{array} 
\right]
=
U_{+2}U_{+1}
\left[
    \begin{array}{c}
        1 \\
        0
    \end{array} 
\right],
\end{equation}
using also Eqs. (\ref{smallU}) and (\ref{equality}).
We obtain
\begin{widetext}
\begin{eqnarray}
A_+(T) &=&  B_+(T) = \frac{\cos{\phi_1}\cos{\phi_2}-w(n_x^2+n_y^2)\sin{\phi_1}\sin{\phi_2}-in_x\cos{\phi_1}\sin{\phi_2}}{\cos{\phi_1}+w(n_y+in_x)\sin{\phi_1}}, \\
A_+(T) &=& B_+(T) = \frac{(n_y-in_x)(\cos{\phi_1}\sin{\phi_2}+w\sin{\phi_1}\cos{\phi_2})+wn_x(n_x+in_y)\sin{\phi_1}\sin{\phi_2}}{\cos{\phi_1}-w(n_y+in_x)\sin{\phi_1}},
\end{eqnarray}
\end{widetext}
and by equating the above fractions and after some manipulation we end up with the relation
\begin{eqnarray}
\label{cond1}
(w\sin{2\phi_1}+n_y\cos{2\phi_1})\sin{\phi_2} &+& \nonumber \\
(wn_y\sin{2\phi_1}-\cos{2\phi_1})\cos{\phi_2} &=& 0.
\end{eqnarray}

The second relation involving the angles $\phi_1, \phi_2$ will be obtained from the condition that the $R$-vector should end up on the $yz$-meridian, which is
\begin{equation}
\label{xzero}
x_-(T)=2\Re{\{A^*_-(T)B_-(T)\}}=0.
\end{equation}
The expressions for $A_-(T), B_-(T)$ are obtained from the corresponding propagation equation and are given in appendix \ref{sec:appendixB}, from which becomes obvious that using Eq. (\ref{xzero}) directly is cumbersome.
The crucial observation which considerably simplifies the analysis is that $\Im{\{A_-(T)\}}=\Im{\{B_-(T)\}}$.
Using this relation in Eq. (\ref{xzero}) we get
\begin{equation}
\label{eqcond}
\Im^2{\{A_-(T)\}}=-\Re{\{A_-(T)\}}\Re{\{B_-(T)\}}.
\end{equation}
It is also $|A_-(T)|^2+|B_-(T)|^2=1$,
which can be written as
\begin{equation}
\label{magn}
\Re^2{\{A_-(T)\}}+\Re^2{\{B_-(T)\}}+2\Im^2{\{A_-(T)\}}=1.
\end{equation}
Using Eq. (\ref{eqcond}) in Eq. (\ref{magn}), we finally end up with
\begin{equation}
\Re{\{A_-(T)\}}-\Re{\{B_-(T)\}}=\pm 1,
\end{equation}
which with the aid of Eqs. (\ref{R_final}) can be expressed as
\begin{eqnarray}
\label{cond2}
(-w\sin{2\phi_1}+n_y\cos{2\phi_1})\sin{\phi_2} &+& \nonumber \\
(wn_y\sin{2\phi_1}+\cos{2\phi_1})\cos{\phi_2}  &=& \pm 1.
\end{eqnarray}

\begin{figure}[t]
 \centering
 \begin{subfigure}[b]{0.4\textwidth}
    \centering\caption{}\includegraphics[width=\linewidth]{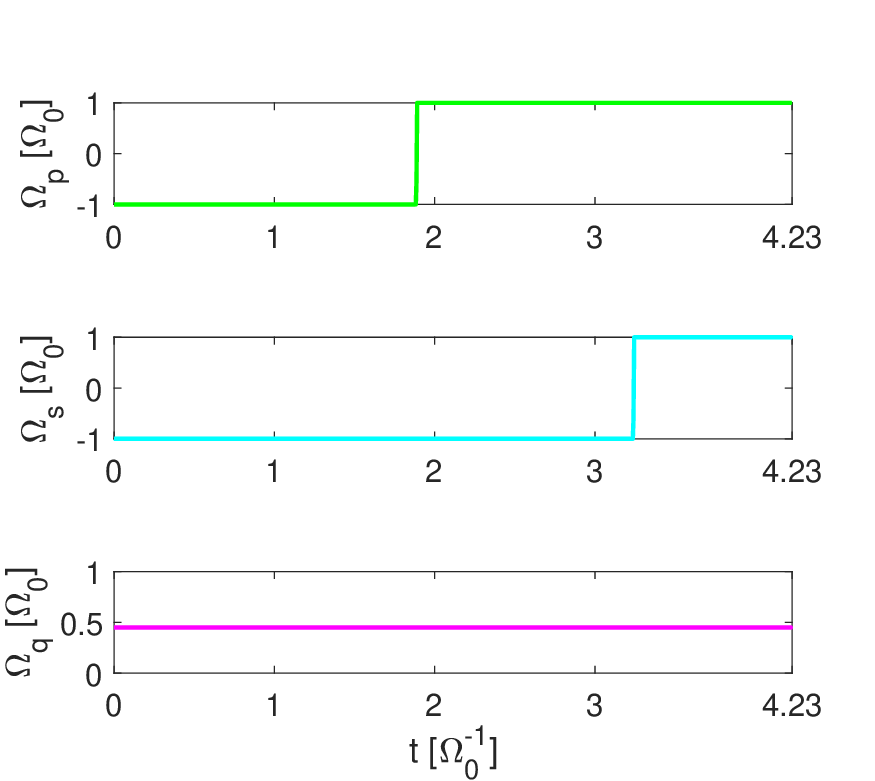}\label{fig:W045c}
\end{subfigure}%
\hspace{.2cm}
\begin{subfigure}[b]{0.4\textwidth}
    \centering\caption{}\includegraphics[width=\linewidth]{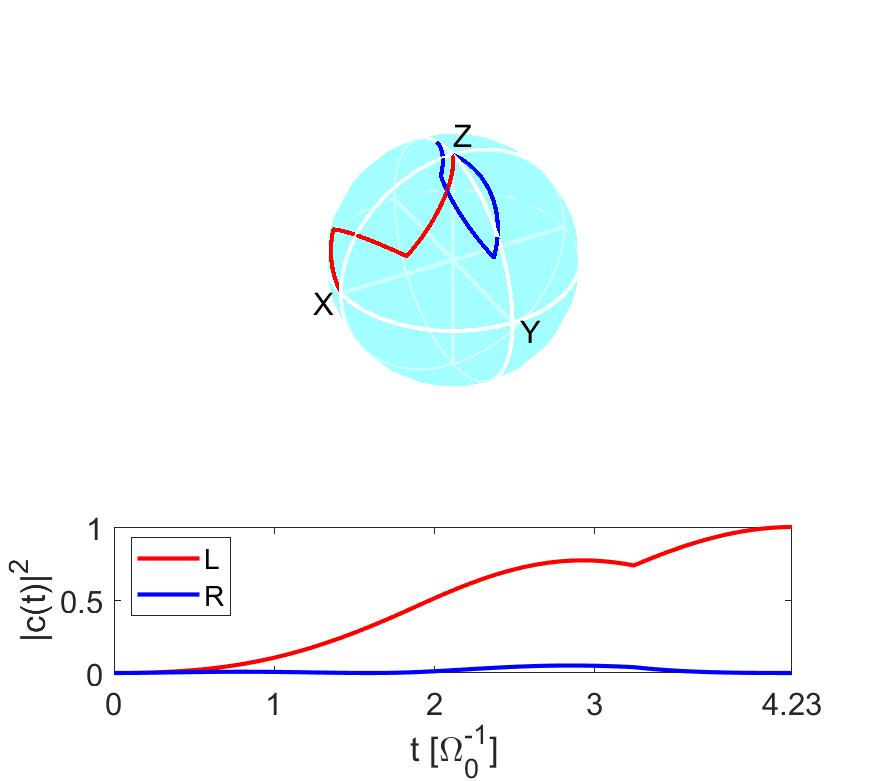}\label{fig:W045}
\end{subfigure}
\caption{Case $\Omega_1/\Omega_0=0.45$, (a) optimal controls, (b) optimal trajectories on the Bloch sphere (upper panel) and time evolution of level $\ket{3}$ populations (lower panel) for the $L$- (red) and $R$- (blue) molecules.}
\label{fig:045}
\end{figure}

\begin{figure}[t]
 \centering
 \begin{subfigure}[b]{0.4\textwidth}
    \centering\caption{}\includegraphics[width=\linewidth]{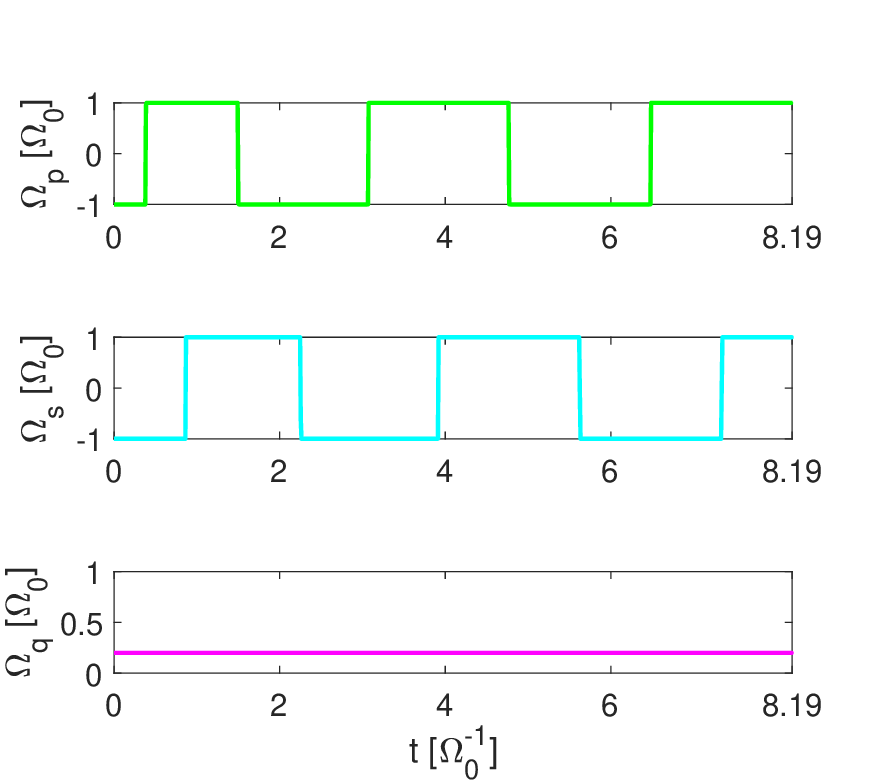}\label{fig:W02c}
\end{subfigure}%
\hspace{.2cm}
\begin{subfigure}[b]{0.4\textwidth}
    \centering\caption{}\includegraphics[width=\linewidth]{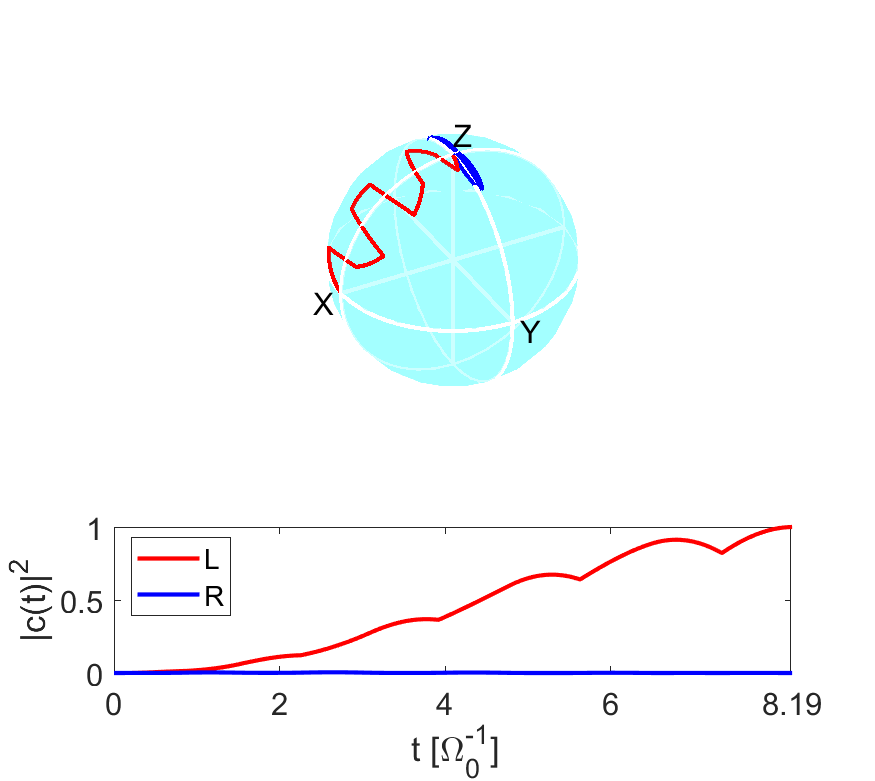}\label{fig:W02}
\end{subfigure}
\caption{Case $\Omega_1/\Omega_0=0.2$, (a) optimal controls, (b) optimal trajectories on the Bloch sphere (upper panel) and time evolution of level $\ket{3}$ populations (lower panel) for the $L$- (red) and $R$- (blue) molecules.}
\label{fig:02}
\end{figure}

Eqs. (\ref{cond1}) and (\ref{cond2}) form a linear system for $\sin{\phi_2}, \cos{\phi_2}$ with solution
\begin{equation}
\label{phi21}
\sin{\phi_2} = \frac{\Delta_s}{\Delta}, \quad \cos{\phi_2} = \frac{\Delta_c}{\Delta},
\end{equation}
where
\begin{subequations}
\label{Dprime}
\begin{eqnarray}
\Delta &=& 2n_y(w^2\sin^2{2\phi_1}+\cos^2{2\phi_1}), \\
\Delta_s &=& \mp (wn_y\sin{2\phi_1}-\cos{2\phi_1}), \\
\Delta_c &=& \pm (w\sin{2\phi_1}+n_y\cos{2\phi_1}).
\end{eqnarray}
\end{subequations}
Using expressions (\ref{Dprime}) in the identity $\sin^2{\phi_2}+\cos^2{\phi_2}=1$, we get after some manipulation
\begin{equation}
\sin{2\phi_1}=\frac{1}{s}\sqrt{\frac{1-s^4}{2}},
\end{equation}
and with the aid of Eq. (\ref{phiprime}) that
\begin{equation}
\label{tauprime}
\tau = \frac{2}{\Omega_0}\frac{1}{\sqrt{s^2+1}}\arcsin{\left[\frac{1}{s}\sqrt{\frac{1-s^4}{2}}\right]}.
\end{equation}
Eq. (\ref{cond1}) can be recast in the following form
\begin{equation}
\tan{(\phi_2+\chi)}=\frac{1}{w}\cot{2\phi_1},
\end{equation}
where 
\begin{equation}
\tan{\chi}=n_y=\frac{s}{\sqrt{s^2+2}},
\end{equation}
leading with the aid of Eq. (\ref{phiprime}) to 
\begin{eqnarray}
\label{t2prime}
T-2\tau &=& \frac{4}{\Omega_0}\frac{1}{\sqrt{s^2+2}}  \\
        &\times & \left[\arctan{\sqrt{\frac{s^4+2s^2-1}{(s^2+2)(1-s^2)}}}-\arctan{\frac{s}{\sqrt{s^2+2}}}\right]. \nonumber
\end{eqnarray}
Note that for $s=1$ it is $\tau=0$ and $T$ attains the value given in Eq. (\ref{const}).
Having found the angles $\phi_1, \phi_2$, we can also calculate the polar angle of the terminal point of the $R$-trajectory on the $yz$-meridian from the relation
\begin{eqnarray}
\label{polar}
\theta(T) &=& \arccos{\left[z_-(T)\right]} \nonumber \\
          &=& \arccos{\left[|A_-(T)|^2-|B_-(T)|^2\right]} \nonumber \\
          &=& \arccos{\left[\Re^2{\{A_-(T)\}}-\Re^2{\{B_-(T)\}}\right]},
\end{eqnarray}
using expressions (\ref{R_final}).

In Fig. \ref{fig:W09c} we display the optimal pulse-sequence for $s=\Omega_1/\Omega_0=0.9$, while in Fig. \ref{fig:W09} the corresponding trajectories of the Bloch vectors (upper panel) and the time evolution of state $\ket{3}$ populations for the the $L$- (red) and $R$- (blue) molecules (lower panel). In Fig. \ref{fig:polar} we plot $\theta(T)$ as a function of $s=\Omega_1/\Omega_0$ for $0.86 \leq s \leq 1$ and observe that when $s=1$ it is $\theta(T)=\pi/2$, so  $\vec{r}_-(T)$ is aligned with the $y$-axis.

For values $s<0.86$, numerical optimal control indicates that the optimal pulse-sequence loses its symmetry and the number of stages increases in general \cite{Boscain06,Evangelakos23}, as shown in Figs. \ref{fig:075}, \ref{fig:045}, \ref{fig:02}, where optimal solutions are obtained with the BOCOP solver for $s=0.75, 0.45, 0.2$, respectively. Note also that for decreasing $s$, the $R$-vector ends up on the $yz$-meridian at points with higher $z$-component. Another interesting observation regarding the optimal sequences is that for the case $s<1$ the $Q$-field is always active and equal to its boundary value, while for $s>1$ the Raman fields show this behavior. It thus appears that in all cases, the fields with the lower control bound are always on and equal to a boundary value. Based on this we can point out that the three-stages symmetric optimal sequences emerge from the constant control solution by removing the $Q$-field in the middle stage for $s>1$, and a different Raman field in the initial and final stages for $s<1$.

We close this section by providing an intuitive understanding of the optimal control values (bang or zero) obtained from the Maximum Principle.
A way to understand chiral resolution using the closed-loop three-level system (\ref{LR_Hamiltonian}) is to interpret it as a quantum interference phenomenon, between the direct pathway connecting states $\ket{1}, \ket{3}$, enabled by the field $\Omega_q$, and the indirect pathway between the same states enabled by the fields $\Omega_{p,s}$. The opposite sign of the $Q$-field for the two enantiomers, combined with the proper choice of the control fields, leads to a constructive interference of state $\ket{3}$ probability amplitudes for the $L$-handed molecules and a destructive interference for the $R$-handed. To achieve this constructive/destructive interference in the minimum possible time, it is intuitively expected to use the boundary values of the fields, since they induce the maximum phase during a given time interval. Now observe that for $s=\Omega_1/\Omega_0=1$ all the corresponding optimal controls assume constant values at their boundaries [Fig. \ref{fig:W1c}]. For $s>1$, if all three controls continue to take constant boundary values, then the direct path would be ahead in phase, i.e. somehow ``faster" than it is needed. The emergence of singular zero control intervals in $\Omega_q$, while $\Omega_{p,s}$ maintain their constant boundary values, assures that the two paths are synchronized to achieve the desired quantum interference [Fig. \ref{fig:W2c}]. For $s<1$, the synchronization is achieved through singular control intervals [Fig. \ref{fig:W09c}], switchings between the boundary values [Figs. \ref{fig:W045c} and \ref{fig:W02c}] or combination of them [Fig. \ref{fig:W075c}] in the controls $\Omega_{p,s}$ enabling the ``faster" indirect path, while $\Omega_q$ maintains a constant boundary value.

\section{Comparison with other pulsed protocols}

\label{sec:comparison}

In this section we compare our time-optimal protocols with other pulse-sequences known to achieve perfect chiral resolution. Below we briefly describe these protocols and obtain the corresponding time needed under constraints (\ref{constraints}), for more details see Refs. \cite{Torosov20a,Torosov20b}.

According to the first ``$\pi/2 - \pi - \pi/2$" protocol, three single pulses with the corresponding areas are sequentially used, each of them applied to one of the transitions of the three-level system, while one of the pulses should have a $\pi/2$ phase relative to the others. To minimize the duration of the protocol, it is necessary to use the maximum amplitude for each pulse, $\Omega_0$ for the $P$- and $S$-pulses and $\Omega_1$ for the $Q$-pulse. We have to distinguish two cases, whether the $Q$-pulse is in the middle or not, so its area is $\pi$ or $\pi/2$, respectively. The corresponding protocol durations, as functions of ratio $s=\Omega_1/\Omega_0$, are
\begin{subequations}
\label{three_pulses}
\begin{eqnarray}
T_{PQS} &=& \frac{\pi}{\Omega_0}\left(1+\frac{1}{s}\right), \\
T_{PSQ} &=& \frac{\pi}{2\Omega_0}\left(3+\frac{1}{s}\right).
\end{eqnarray}
\end{subequations}

We next continue with the ``single-Raman-single" pulse-sequence \cite{Torosov20b}, which also consists of three stages. The sequence starts with a $Q$-pulse with area $\pi/2$, followed by a Raman interaction consisting of two simultaneous $P$- and $S$-pulses, where both have area $\pi/\sqrt{2}$ while the $S$-pulse has additionally a relative phase $\pi/2$. The protocol ends as it started, with a $Q$-pulse of area $\pi/2$ applied at the final stage. The total duration of this protocol is 
\begin{equation}
\label{sRs}
T_{Q(PS)Q} = \frac{\pi}{\Omega_0}\left(\frac{1}{\sqrt{2}}+\frac{1}{s}\right).
\end{equation}

The final protocol is the ``Raman-single" pulse-sequence, which has only two stages. The Raman stage consists of two simultaneous $S$- and $P$-pulses with relative phase $\pi/2$ and areas $\pi\sqrt{2-\sqrt{2}}$ and $\pi\sqrt{2+\sqrt{2}}$, respectively, while during the second stage a single $Q$-pulse with $\pi/2$ area is applied. For the implementation of the protocol under constraints (\ref{constraints}), during the Raman stage we set $\Omega_p = \Omega_0$, thus $\Omega_s=\sqrt{\frac{2-\sqrt{2}}{2+\sqrt{2}}}\Omega_0 = (\sqrt{2}-1)\Omega_0$. The total duration of the protocol is
\begin{equation}
\label{Rs}
T_{(PS)Q} = \frac{\pi}{\Omega_0}\left(\sqrt{2+\sqrt{2}}+\frac{1}{2s}\right).
\end{equation}

\begin{figure}[t]
 \centering
\includegraphics[width=\linewidth]{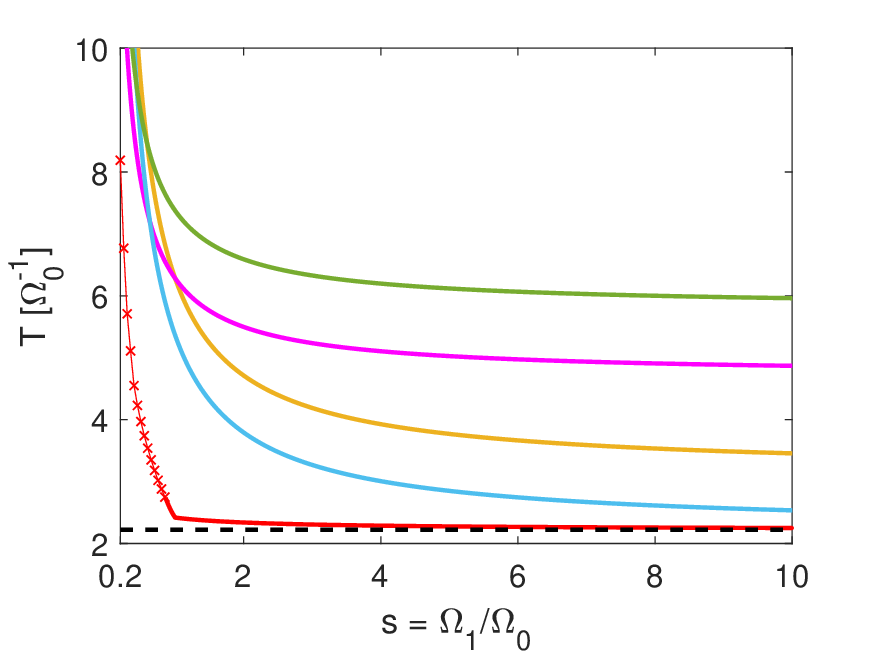}
\caption{Duration of various pulsed protocols for perfect chiral resolution versus the maximum amplitudes ratio $s=\Omega_1/\Omega_0$: time-optimal (red), ``$\pi/2 - \pi - \pi/2$" (yellow for $Q$-pulse in the middle, magenta for $Q$-pulse at the initial or final stages), ``single-Raman-single" (blue) and ``Raman-single" (green).}
\label{fig:comparison}
\end{figure}

In Fig. \ref{fig:comparison} we display the time required by the various pulsed protocols for perfect chiral resolution versus the maximum amplitudes ratio $s=\Omega_1/\Omega_0$: time-optimal (red), ``$\pi/2 - \pi - \pi/2$" (yellow for $Q$-pulse in the middle, magenta for $Q$-pulse at the initial or final stages), ``single-Raman-single" (blue) and ``Raman-single" (green). The horizontal black dashed line indicates the limit (\ref{glimit}). For the optimal protocol, we plot with red solid line the analytical values obtained for $s\geq 0.86$ in Secs. \ref{sec:greater} and \ref{sec:lower}, while with red crosses the minimum-time values found using the numerical optimal control solver BOCOP for $0.2\leq s \leq 0.85$ with step $\delta s = 0.05$. As expected, the optimal solution achieves perfect chiral resolution faster than the other pulsed protocols and approaches the limit (\ref{glimit}) for $s\rightarrow \infty$. The better performance of the optimal protocol is attributed to that during its application more controls are simultaneously active than in the other protocols. Another interesting observation regarding the optimal protocol is the dramatic reduction in duration $T$ when the bound on the $Q$-field approaches the bound on the Raman fields from lower values ($\Omega_1\rightarrow\Omega_0^-$) so the constant control solution emerges.
Among the other protocols, only the ``single-Raman-single" tends to the same limiting value (\ref{glimit}) for $s\rightarrow \infty$, while the rest tend to higher values in the same limit, see also Eqs. (\ref{three_pulses}), (\ref{sRs}) and (\ref{Rs}). Regarding the ``$\pi/2 - \pi - \pi/2$" protocols, the faster for each value of $s$ is the one where the larger $\pi$-pulse is executed with the control with the higher maximum amplitude. Thus, the protocol with the $Q$-pulse in the middle is faster when $s>1$  and slower for $s<1$. The bad performance of the ``Raman-single" protocol for large $s$ is due to the fact that the large area $P$-pulse is executed with the smaller maximum amplitude $\Omega_0$. As we point out at the end of Sec. \ref{sec:lower}, controls $\Omega_{p,s}$ and $\Omega_q$ enable the indirect and direct paths, respectively, when chiral resolution is seen as a quantum interference phenomenon, thus the different sensitivity to them can be understood from this point of view.




\section{Conclusion}
\label{sec:conclusion}

We formulated the problem of obtaining in minimum-time perfect chiral resolution with bounded control fields, as an optimal control problem on two non-interacting spins-$1/2$, assuming a common control bound for the two Raman fields (pump and Stokes) and a different bound for the field connecting directly the two lower-energy states. We used optimal control theory to show 
that the only values permitted for the optimal fields are the control bounds and zero, with the latter corresponding to singular control.
Subsequently, with the aid of numerical optimal control and intuition, we identified some three-stage symmetric optimal pulse-sequences, for relatively larger values of the ratio between the two control bounds, and derived analytical expressions for the timings of the constituent pulses as functions of this ratio. For lower values of the bounds ratio, numerical optimal control suggested that the optimal pulse-sequences are no longer symmetric and include more stages in general. We compared the analytical/numerical optimal protocol with other pulsed protocols under the same control constraints, and found it to achieve faster perfect chiral resolution, primarily because of the simultaneous action of the control fields. The present study is expected to find application in the various fields of natural sciences where enantiomer separation plays a crucial role. 

We close this article with some observations which pave the way for future work. The closed-loop three-level system considered here, from which arises the system of two non-interacting spins, is used in most of the references cited in the introduction, while in Refs. \cite{Kral03,Thanopulos03,Gerbasi04,Thanopulos05} is employed to model specific molecules. Since the aim of the current theoretical work is to derive the minimum-time pulses achieving perfect chiral resolution, these pulses are generally expected to be more prone to experimental errors and variations in Hamiltonian parameters than for example longer composite pulses specifically designed to alleviate such imperfections, as in Ref. \cite{Torosov20a}.
Regarding the bang-bang nature of the optimal solution, note that such controls are routinely considered even for more complex Hamiltonians in the context of Quantum Approximate Optimization Algorithm, see for example Ref. \cite{Brady21}. Nevertheless, for possible experimental implementation of the bang-bang pulses in a real system, one may use numerical optimization to obtain pulses with short but finite rise times. Regarding noise sensitivity, it depends on the type of noise. For example, in the case of colored noise, if the inverse duration of the fast chiral resolution achieved with the minimum-time pulses corresponds to frequencies outside of the range where the noise is substantial, then the pulse-sequence is immune to this type of noise. For a real molecular system with known noise sources one can actually utilize numerical optimization to derive optimal pulses with respect to system imperfections and noise. The duration of these pulses should be longer than the minimum-time pulses and long enough to ensure the desired robustness. The above considerations motivate our future work, which will be to test the proposed theoretical method using a real molecule and adapt it to possible limitations which may arise, probably following this hybrid approach with a trade-off between speed and robustness.

\appendix

\section{Expressions for $A_-(T), B_-(T)$ when $0.86\leq s < 1$}

\label{sec:appendixB}

Using the propagation equation
\begin{equation}
\left[
    \begin{array}{c}
        A_-(T) \\
        B_-(T)
    \end{array} 
\right]
=
U_{-3}U_{-2}U_{-1}
\left[
    \begin{array}{c}
        1 \\
        0
    \end{array} 
\right],
\end{equation}
we find
\begin{subequations}
\label{R_final}
\begin{eqnarray}
\Re{\{A_-(T)\}} &=& \cos^2{\phi_1}\cos{\phi_2}-\frac{2}{w}\cos{\phi_1}\sin{\phi_1}\sin{\phi_2} \nonumber \\
            &-& w^2n_y^2\sin^2{\phi_1}\cos{\phi_2}+w^2n_x^2n_y\sin^2{\phi_1}\sin{\phi_2}, \nonumber \\
            \\
\Im{\{A_-(T)\}} &=& -n_x\cos^2{\phi_1}\sin{\phi_2}-wn_x\cos{\phi_1}\sin{\phi_1}\cos{\phi_2} \nonumber \\
            &+& w^2n_x^3\sin^2{\phi_1}\sin{\phi_2}-wn_xn_y\cos{\phi_1}\sin{\phi_1}\sin{\phi_2}  \nonumber \\
            &-& w^2n_xn_y\sin^2{\phi_1}\cos{\phi_2},  \\
\Re{\{B_-(T)\}} &=& -n_y\cos^2{\phi_1}\sin{\phi_2}-2wn_y\cos{\phi_1}\sin{\phi_1}\cos{\phi_2} \nonumber \\
            &+& 2wn_x^2\cos{\phi_1}\sin{\phi_1}\sin{\phi_2}+w^2n_y\sin^2{\phi_1}\sin{\phi_2} \nonumber \\ 
            &+& w^2n_x^2\sin^2{\phi_1}\cos{\phi_2}, \\
\Im{\{B_-(T)\}} &=& \Im{\{A_-(T)\}}.           
\end{eqnarray}
\end{subequations}

\begin{acknowledgements}
None
\end{acknowledgements}



\bibliographystyle{apsrev4-2}
\bibliography{main}

\vspace*{1. cm}

\end{document}